\documentclass[11pt]{article}
\usepackage{amssymb,amsfonts,amsmath,mathrsfs}
\newcommand{\ot}[0]{\otimes}

\newcommand{\ket}[1]{|#1\rangle}
\newcommand{\bra}[1]{\langle#1|}

\newcommand{\proj}[1]{\ket{#1}\!\bra{#1}}

\def\LRA{\mathop{-\!\!\!-\!\!\!
\longrightarrow}\nolimits}

\newcommand{\mac}[1]{\mathsf{#1}}
\newcommand{\stan}[0]{\varrho_{ABA'B'}}
\newcommand{\ten}[1]{\mathcal{#1}}
\newcommand{\Def}[2]{\noindent{\bf Definition} {\bf #1.} {\it #2}}
\newcommand{\Th}[2]{\noindent{\bf Theorem} {\bf #1.} {\it #2}}

\newcommand{\s}[1]{\mathscr{#1}}

\newcommand{\bb}[1]{\mathbb{#1}}
\newcommand{\be}[0]{\begin{equation}}
\newcommand{\ee}[0]{\end{equation}}
\newcommand{\bea}[0]{\begin{eqnarray}}
\newcommand{\eea}[0]{\end{eqnarray}}

\def\LRA{\mathop{-\!\!\!-\!\!\!
\longrightarrow}\nolimits}
\newcommand{\Tr}[0]{\mathrm{Tr}}
\newcommand{\pbit}{\gamma_{ABA'B'}^{(2)}}
\newcommand{\pdit}{\gamma_{ABA'B'}^{(d)}}
\newcommand{\norsl}[1]{\left|\left|#1\right|\right|_{\Tr}}
\newcommand{\baza}{\s{B}_{AB}}
\newcommand{\maks}{\ket{\Psi_{+}^{(d)}}}
\newcommand{\mapr}{\ten{P}_{+}^{(d)}}
\begin{document}
\title{\LARGE On quantum cryptography with bipartite bound entangled states}

\author{Pawe\l{} Horodecki\footnote{e-mail:~\texttt{pawel@mif.pg.gda.pl}}~ and Remigiusz Augusiak \\
 {\it {\small Faculty of Applied Physics and
Mathematics}}, {\it {\small Gda\'nsk University of Technology}},\\
{\it {\small Narutowicza 11/12, PL 80-952 Gda\'nsk, Poland}} }
\date{}

\maketitle

\abstract{Recently the explicit applicability of bound entanglement in quantum cryptography has been shown.
In this paper some of recent results respecting this
topic are reviewed. In particular relevant notions and definitions are reminded. The new construction of bound entangled states containing secure correlations is presented. It
provides low dimensional $6 \otimes 6$ bound entangled states with nonzero distillable key.}
\section{Introduction}
The explicit application of quantum entanglement in quantum
information theory was the cryptographic protocol by Ekert
\cite{Ekert91}. The essential point of the protocol (cf. further
modification \cite{Bennett92}) was the entanglement monogamy
principle (see \cite{Terhal}) which says that if the two particles
are maximally entangled with each other then they are completely
unentangled with any other (third) party. Hence, results of any
correlation measurements on both particles must be completely safe
from cryptographic point of view, as they are uncorrelated form
results of any other measurement performed on the rest of the
world. This point was further exploited in a nice application
\cite{Deutsch96} of entanglement distillation \cite{puryfikacja} (cf. \cite{puryfikacja1}).
The idea of Ref. \cite{Deutsch96} is called quantum privacy
amplification (QPA). Given stationary source of pure states
$\ket{\Psi_{ABE}}$  describing Alice, Bob (which are cooperating)
and Eve (eavesdropper) quantum correlations protocol QPA is
focused on distilling maximally entangled states

\begin{equation}\label{max}
\maks=\frac{1}{\sqrt{d}}\sum_{i=0}^{d-1} \ket{i}\ot\ket{i}
\end{equation}
from quantum entangled states
\begin{equation}
\varrho_{AB}=\Tr_{E}\proj{\Psi_{ABE}}.
\end{equation}
The distillation protocol uses local operations and classical
communication (LOCC) which in presence of Eve are usually called
local operations and public communication (LOPC). Once Alice and
Bob distill maximally correlated states $\maks$ by entanglement
monogamy they share $\log d$ bits of classical secure bits. This
can be done by performing local measurements on the state
(\ref{max}) in standard bases $\{|i\rangle_{A} \}_{i=0}^{d-1}$ and
$\{|i\rangle_{B} \}_{i=0}^{d-1}$.

However, since 1998 it has been known that there is quantum
entanglement called bound entanglement that can not be distilled
to a pure form \cite{bound}. For a long time there was a common
belief that distillation of secure key from quantum state is
possible only when QPA is. In other words, that bound entanglement is useless for quantum cryptography. In fact the results of extensive analysis of two qubit case \cite{Gisin}
naturally suggested equivalence of entanglement distillation
protocols and secure key distillation.

Surprisingly it is not true, as it has been shown in papers
\cite{KH2,KH}. Before we shall recall main observations of the
latter, let us point out the key ingredient of their reasoning.
Namely, why QPA might be not necessary for distilling secure key?
In fact, if Alice and Bob share maximally entangled state
(\ref{max}) they in a sense have much stronger security than they
need. In fact they will get secure correlations if they measure
the state in {\it any} pair of bases of the form:
$\{\ket{e^{(A)}_{i}}=U|i\rangle_{A} \}_{i=0}^{d-1}$ and
$\{\ket{e^{(B)}_{i}}=U^{*}|i\rangle_{B} \}_{i=0}^{d-1}$, where $U$
stands for arbitrary unitary operation. It is crucial to
understand that in quantum cryptography it would be enough to have
{\it a single basis}, measurement in which could give secure
correlations. To be more specific, the security requirement is to
get (via local measurement of Alice and Bob in one basis) the
state that is product with Eve's degrees of freedom.

This leads us to the notion of ccq state \cite{Devetak}, i.e.,
tripartite state of Alice, Bob and Eve that is, after local von
Neumann measurements, classical on Alice and Bob parts and quantum
on Eve part. With this notion one can summarize the idea of
\cite{KH2,KH} as follows: Alice and Bob should proceed to distill
such a state $\pbit$ that, after considering its purification
$\ket{\Psi_{ABA'B'E}}$ and performing a local measurements in a
standard product basis on its $AB$ part the resulting ccq state
$\varrho_{ABE}$ is (i) product with respect to the division $AB|E$
(ii) contains maximal classical correlations between Alice and
Bob.

In this work we shall briefly describe main mathematical elements
of the construction \cite{KH2,KH}, recall the idea of one-way
distillation provided in \cite{KH3} and provide a new construction
of bound entangled states with secure quantum key.

%%%%%%%%%%%%%%%%%%%%%%%%%%%%%%%%%%%%%%%%%%%%%%%%%%%%%%%%%%%%%%%%%
%%%%%%%%%%%%%%%%%%%%%%%%%%%%%%%%%%%%%%%%%%%%%%%%%%%%%%%%%%%%%%%%%
\section{Main notions of general secure key distillation scheme}
%%%%%%%%%%%%%%%%%%%%%%%%%%%%%%%%%%%%%%%%%%%%%%%%%%%%%%%%%%%%%%%%%
%%%%%%%%%%%%%%%%%%%%%%%%%%%%%%%%%%%%%%%%%%%%%%%%%%%%%%%%%%%%%%%%%

%%%CHANGE
Here we shall remind and discuss main ideas and notions
%%%%ENDofCHANGE
introduced in Refs. \cite{KH2,KH}.

Assume that Alice and Bob wish to communicate but without
participation of an eavesdropper Eve. To this aim, as a source of
quantum correlations, they use a quantum state $\stan$. Here,
subsystem $AA'$ ($BB'$) belongs to Alice (Bob). Moreover,
following \cite{KH2}, subsystem $AB$ ($A'B'$) is called the {\it
key part (shield part)} of the given state. To make the
considerations more formal, each subsystem of $\stan$ shall be
represented by respective Hilbert space, i.e., Alice's subsystems
by $\s{H}_{A}$ and $\s{H}_{A'}$ and Bob's by $\s{H}_{B}$ and
$\s{H}_{B'}$. Hence, the key part of $\stan$ is defined on
$\s{H}=\s{H}_{A}\ot \s{H}_{B}$ and the shield part on
$\s{H}'=\s{H}_{A'}\ot \s{H}_{B'}$.
%More formally, we ascribe to each subsystem we ascribe Hilbert
%spaces, respectively, $\s{H}_{A}$ and $\s{H}_{A'}$ to Alice
%subsystems and $\s{H}_{B}$ and $\s{H}_{B'}$ to Bob.
Hereafter we shall be assuming that
$\dim\s{H}_{A}=\dim\s{H}_{B}=d$, $\dim\s{H}_{A'}=d_{A'}$, and
$\dim\s{H}_{B'}=d_{B'}$.

Firstly, let us remind the notion of ccq state. To this aim let us
introduce a product basis defined on Hilbert space $\s{H}_{A}\ot
\s{H}_{B}\sim\mathbb{C}^{d}\ot\mathbb{C}^{d}$, i.e.,
\begin{equation}
\baza=\left\{\ket{e_{i}^{(A)}}\ot\ket{e_{j}^{(B)}}\right\}_{i,j=0,\ldots,d-1},
\end{equation}
where $\{\ket{e_{i}^{(A)}}\}_{i=0,\ldots,d-1}$ and
$\{\ket{e_{i}^{(B)}}\}_{i=0,\ldots,d-1}$ are arbitrary bases
spanning the Hilbert spaces, respectively, $\s{H}_{A}$ and
$\s{H}_{B}$. Of course, these bases may be chosen to be standard
and therefore $\baza$ is the
standard basis in $\bb{C}^{d}\ot\bb{C}^{d}$. Then we have the following \\

\Def{1}{(ccq state) We call the state $\tilde{\varrho}_{ABE}$ a
{\bf ccq state} of initial state $\stan$ with respect to the basis
$\baza$ if $\tilde{\varrho}_{ABE}$ is a result of measurement of
$\varrho_{ABE}=\Tr_{A'B'}\proj{\Psi_{ABA'B'E}}$ in the product
basis $\baza$. Here $\ket{\Psi_{ABA'B'E}} $ is a purification of
$\stan$.\\}

As an illustrative example let us consider a density matrix
$\eta_{ABA'B'}$ acting on $({\bb C}^{2} )^{\otimes 4}$ and given
by $\eta_{ABA'B'}=p \proj{0111}+(1-p)\proj{1000}$. As one may
easily verify its standard purification takes the form
$\ket{\Psi_{ABA'B'E}}= \sqrt{p}\ket{01110}+\sqrt{1-p}\ket{10001}$.
Therefore, the ccq state of $\eta_{ABA'B'}$ with respect to
standard basis $\baza^{(\mathrm{st})}\equiv \{
|i\rangle_{A}|j\rangle_{B} \}_{i,j=0}^{1}$ is
$\tilde{\eta}_{ABE}=p\proj{010}+(1-p)\proj{101}$.

Now, one can ask when a given state is said to be secure.
As an answer one gives the following (see \cite{KH})  \\

\Def{2}{(security) We call the state $\stan$ {\bf secure} with
respect to the basis $\baza$ if its ccq state is of the form}
\begin{equation}
\tilde{\varrho}_{ABE}=\left[\sum_{i,j=0}^{d-1}p_{ij}\proj{e_{i}^{(A)}e_{j}^{(B)}}\right]\ot\varrho_{E}.
\end{equation}
The security of such a state follows from the fact that Eve is
completely uncorrelated from distribution represented by $AB$
system after Alice and Bob measurement. Note that if the
distribution $p_{ij}$ is homogenous, i.e.,
$p_{ij}=1/d;(i,j=0,\ldots,d-1)$, then we say that $\stan$ has a
{\bf $\baza$-key}.

A very important ingredient of construction discussed here is a
special class of controlled
unitary operations (see \cite{KH2,KH}) that we recall by the following  \\

\Def{3}{(twisting) Let $\mac{U}_{ij}^{(A'B')}$ be certain
unitary operations acting on subsystem $A'B'$. For a given basis $\baza$
we call the operation}
\begin{equation}\label{twisting}
\mac{U}=\sum_{i,j=0}^{d-1}\proj{e_{i}^{(A)}e_{j}^{(B)}}\ot
\mac{U}_{ij}^{(A'B')}
\end{equation}
{\it {\bf $\baza$-twisting} or shortly, {\bf twisting}.\\}

The importance of such a class of operations follows from the fact
that applied to a given state $\stan$, $\mac{U}$ {\it do not}
change its ccq state. It means that if we take $\stan$ and
$\sigma_{ABA'B'}=\mac{U}\stan\mac{U}^{\dagger}$, their ccq states
are exactly the same, i.e.,
$\tilde{\varrho}_{ABE}=\tilde{\sigma}_{ABE}$.

Now we define the central notion of the generalised approach
provided in \cite{KH2,KH}. This is the notion of private state
that has $\log d$ bits of secure key encoded in its  $AB$ part of
$d \otimes d$ type).\\

\Def{4}{(pdit) Let $\stan$ is a density operator acting on the
Hilbert space $\s{H}\ot\s{H}'$ and $\varrho_{A'B'}$ is a density
matrix acting on $\s{H}'$. Let
$\mac{U}_{i}^{(A'B')}\;(i=0,\ldots,d-1)$ be certain unitary
operations acting on $A'B'$ system. Then we call the state $\stan$
{\bf private state} or {\bf pdit} with respect to the basis
$\baza$ if it is of the form}
\begin{equation}
\stan=\frac{1}{d}\sum_{i,j=0}^{d-1}\ket{e_{i}^{(A)}e_{i}^{(B)}}\bra{e_{j}^{(A)}e_{j}^{(B)}}\ot
\mac{U}_{i}^{(A'B')}\varrho_{A'B'}\mac{U}_{j}^{(A'B')\dagger}.
\end{equation}
Hereafter, as usual, we shall denote the private dits by $\pdit$.
In the case when the dimension of the key part is $d=2$ on each
side, we have to do with {\bf private bit} or {\bf pbit}.

Now it is important to pose the question: {\it  What is the
essential feature that allows private bit to be truly private?}
Namely, a detailed analysis shows \cite{KH2,KH} that ccq state of
private bit $\pdit$ is the same as the ccq state of the following
state (called {\bf basic pdit})
\begin{equation}\label{basic pdit}
\mapr \otimes \sigma_{A'B'},
\end{equation}
where $\sigma_{A'B'}=\Tr_{AB}\pdit$ and $\mapr$ is a projector
onto $\maks$. This is because it is always possible for a given
pbit to find such a twisting under which $\pbit$ transforms it to
(\ref{basic pdit}) and conversely. Thus, after performing
measurement in the basis $\baza$ the physical system $ABE$ is in
the same state irrespective of whether before Alice and Bob state
was (\ref{basic pdit}) or just a pdit $\pdit$. Hence the security
with respect to the measurement in {\it that particular basis} is
the same as if Alice and Bob really shared maximal entangled state
$\mapr$! This is the key observation for understanding the essence
of private dit.

We conclude the preliminary section racalling the definition of
distillable key \cite{KH2,KH} and related theorem.\\

\Def{5}{(distillable key) Let $\sigma_{AB}$ be a density matrix
acting on $\bb{C}^{d_{A}}\ot \bb{C}^{d_{B}}$ and let
$\mathscr{P}_{n}$ be a sequence of LOCC operations such that
$\mathscr{P}_{n}(\sigma_{AB}^{\ot n})=\Sigma_{n}$ , where
$\Sigma_{n}$ is defined on
$\bb{C}^{d_{n}}\ot\bb{C}^{d_{n}}\ot\s{H}_{A'}\ot\s{H}_{B'}$. The
set $\mathscr{P}=\{\mathscr{P}_{n}\}_{n=1}^{\infty}$ is said to be
a {\bf pdit distillation protocol} of a given state $\sigma_{AB}$
if the following relation }
\begin{equation}
\lim_{n\to\infty}\norsl{\Sigma_{n}-\gamma_{ABA'B'}^{(d_{n})}}=0.
\end{equation}
{\it The rate of this protocol is defined as}
\begin{equation}
\ten{R}(\mathscr{P})=\limsup_{n,d_{n}\to\infty}\frac{\log
d_{n}}{n}
\end{equation}
{\it and distillable key of $\sigma_{AB}$ as}
\begin{equation}
K_{D}(\sigma_{AB})=\sup_{\mathscr{P}}\ten{R}(\mathscr{P}).
\end{equation}

There is a problem however, since the above definition has a
complicated form. It is hard to see whether given state fulfills
the above condition or not. Fortunately, one can simplify the task
providing necessary and sufficient conditions for nonzero
distillable key that are more operational then the definition
itself. Here we provide a summary of the conditions proven in
\cite{KH2,KH}, which are enough to analyse cryptographic
usefulness of many quantum states:\\

\Th{1}{The following three conditions are equivalent:

(i) finite number of copies of state $\stan$ can be transformed
with some LOCC protocol into the state $\sigma_{ABA'B'}$
arbitrarily close in the trace norm to certain pbit $\pbit$

(ii)  finite number of copies of state $\stan$ can be transformed
with some LOCC protocol into the state $\sigma_{ABA'B'}$ with
trace norm of the element $\mac{A}_{00,11}^{(A'B')}$ arbitrarily
close to $1/2$, where the element is defined by the
representation:
\begin{equation}\label{Th11}
\stan=
\sum_{i,j,k,l=0}^{1}\ket{e_{i}^{(A)}e_{j}^{(B)}}\bra{e_{k}^{(A)}e_{l}^{(B)}}
%\ot
%\mac{U}_{i}^{(A'B')}\varrho_{A'B'}\mac{U}_{j}^{(A'B')\dagger
%\sum_{i,j,k,l=0}^{1}\ket{ij}\bra{kl}
\ot \mac{A}_{ij,kl}^{(A'B')}.
%\label{element}
\end{equation}

(iii) the state $\varrho$ has nonzero distillable key, i.e., one
has $K_{D}(\varrho)>0$.

Moreover any convergence of $||\mac{A}_{00,11}^{(A'B')}||_{\Tr}$
to $1/2$ from (ii) during a given protocol is equivalent to a
convergence of the state to a certain pbit during that
protocol.}\\

%The above theorem is also true if instead of pbit we consider pdit
%$\pdit$. Then the trace norm of an element
%$\mac{A}^{(A'B')}_{00,d-1,d-1}$ should converge to $1/d$.

Especially the condition (ii) serves as a useful
criterion which allows to adjudge an applicability
of a given state to quantum cryptography. In next section
we shall illustrate its power.

\section{New class of bound
entangled states with secure quantum key}
%

%

%%%%CHANGE
In this section we present the main result.
%%%%ENDofCHANGE
We provide a
construction of a state that is useful for quantum cryptography
simultaneously being bound entangled. Note that the construction,
however based on that presented in \cite{KH3}, is different in
details from known so far \cite{KH2,KH,KH3} and sheds some light
on the still unexplored domain of bound entanglement. Hereafter it
is assumed that $d=2$, i.e., the key part of $\stan$ consists of
qubits and that $\dim\s{H}_{A'}=\dim\s{H}_{B'}=D$, which allows us
to write $\s{H}'=\bb{C}^{D}\ot\bb{C}^{D}$. We may also assume for
simplicity that $\baza$ is standard basis in
$\bb{C}^{2}\ot\bb{C}^{2}$.

\subsection{The construction}

To obtain a better insight into the construction we begin our
considerations from an illustrative example with the shield part
of dimension $D=3$ on each side. This with the assumption that
$d=2$ makes the considered state to be of dimension $6$ in each
side. Finally, we show that the construction is possible for
arbitrary $D\ge 3$.

\subsubsection{The $6 \otimes 6$ case}
%Here the proposed states will have the shield
%part of dimension $D=3$ in each side. Since the secure key
%part will have dimension $2$ on each side
%whis will give totally local dimension $6$
%of the constructed state.
%%%CHANGE
At the very beginning suppose that Alice and Bob possess
the following state (cf. \cite{KH3})
%%%ENDofCHANGE
%%%%
%
\begin{equation}
\varrho_{ABA'B'}=\frac{11}{40} \left[
\begin{array}{cccc}
|X_{3}| & 0 & 0 & X_{3}\\
0 & \left|X^{T_{B'}}_{3}\right|^{T_{B'}} & 0 & 0\\
0 & 0 & \left|X^{T_{B'}}_{3}\right|^{T_{B'}} & 0\\
X_{3} & 0 & 0 & |X_{3}|
\end{array}
\right]
\end{equation}
where the superscript $T_{B'}$ denotes the partial transposition
with respect to the system $B'$ and $X_{3}$ is a symmetric
$9\times 9$ matrix of the form
\begin{equation}
X_{3}=\frac{1}{11}\left[
\begin{array}{ccccccccc}
-1 & 0 & 0 & 0 & 1 & 0 & 0 & 0 & 1\\
0 & 1 & 0 & 0 & 0 & 0 & 0 & 0 & 0\\
0 & 0 & 1 & 0 & 0 & 0 & 0 & 0 & 0\\
0 & 0 & 0 & 1 & 0 & 0 & 0 & 0 & 0\\
1 & 0 & 0 & 0 & -1 & 0 & 0 & 0 & 1\\
0 & 0 & 0 & 0 & 0 & 1 & 0 & 0 & 0\\
0 & 0 & 0 & 0 & 0 & 0 & 1 & 0 & 0\\
0 & 0 & 0 & 0 & 0 & 0 & 0 & 1 & 0\\
1 & 0 & 0 & 0 & 1 & 0 & 0 & 0 & -1
\end{array}
\right].
\end{equation}
which is defined on the Hilbert space $\s{H}_{A'} \otimes
\s{H}_{B'}= \bb{C}^{3} \otimes \bb{C}^{3}$ . In order to show that
the matrix $\stan$ is Hermitian and nonnegative, i.e., represents
a quantum state, let us observe that $X$ may be decomposed
as\linebreak
$X_{3}=(1/11)(\ten{P}_{+}^{(3)}-2\ten{P}^{(3)}+\ten{Q}^{(3)})$.
Here $\ten{P}_{+}^{(3)}$ is a projector onto the maximally
entangled state
$\ket{\Psi_{+}^{(3)}}=(1/\sqrt{3})\sum_{i=0}^{2}\ket{i}\ket{i}$
belonging to $\mathbb{C}^{3}\ot \mathbb{C}^{3}$ and
$\ten{P}^{(3)}$, and $\ten{Q}^{(3)}$ are projectors given by
relations
\begin{equation}
%\ten{P}^{(3)}=\frac{1}{3}\left(2\sum_{i=0}^{2}\proj{ii}-\sum_{\substack{i,j=0\\i\neq
%j}}^{2}\ket{ii}\bra{jj}\right),
\ten{P}^{(3)}=\sum_{i=0}^{2}\proj{ii}-\ten{P}^{(3)}_{+}, \qquad
\ten{Q}^{(3)}=\ten{I}_{9}-\sum_{i=0}^{2}\proj{ii},
\end{equation}
where $\ten{I}_{9}$ stands for an identity acting on the Hilbert
space $\mathbb{C}^{3}\ot \mathbb{C}^{3}$. These above projectors
are orthogonal and therefore one obtains
\begin{equation}
|X_{3}|=\frac{1}{11}\left[\ten{P}_{+}^{(3)}+2\ten{P}^{(3)}+\ten{Q}^{(3)}\right]=\frac{1}{11}\left[
\begin{array}{ccccccccc}
\frac{5}{3} & 0 & 0 & 0 & -\frac{1}{3} & 0 & 0 & 0 & -\frac{1}{3}\\
0 & 1 & 0 & 0 & 0 & 0 & 0 & 0 & 0\\
0 & 0 & 1 & 0 & 0 & 0 & 0 & 0 & 0\\
0 & 0 & 0 & 1 & 0 & 0 & 0 & 0 & 0\\
-\frac{1}{3} & 0 & 0 & 0 &\frac{5}{3}  & 0 & 0 & 0 & -\frac{1}{3}\\
0 & 0 & 0 & 0 & 0 & 1 & 0 & 0 & 0\\
0 & 0 & 0 & 0 & 0 & 0 & 1 & 0 & 0\\
0 & 0 & 0 & 0 & 0 & 0 & 0 & 1 & 0\\
-\frac{1}{3} & 0 & 0 & 0 & -\frac{1}{3} & 0 & 0 & 0 & \frac{5}{3}\\
\end{array}
\right].
\end{equation}
From the above equation one infers two facts, first that the trace
norm\footnote{For an arbitrary matrix $\Xi$ the trace norm is
defined by relation $||\Xi||_{\rm Tr}={\rm
Tr}\sqrt{\Xi^{\dagger}\Xi}$.} of $X_{3}$ is $||X_{3}||_{\rm Tr}=1$
and the second that $|X_{3}|^{T_{B'}}\ge 0$. Moreover, the matrix
$X_{3}$ partially transposed with respect to the subsystem $B'$
may be written in the form
$X^{T_{B'}}_{3}=(1/11)\left[2\ten{S}^{(3)}-\left(\ten{I}_{9}-\ten{Q}^{(3)}\right)\right]$,
where $\ten{S}^{(3)}$ is a projector given by
\begin{equation}\label{S3}
\ten{S}^{(3)}=\frac{1}{2}\left[\ten{I}_{9}+\ten{V}^{(3)}-2\sum_{i=0}^{2}\proj{ii}\right]
\end{equation}
with $\ten{V}^{(3)}$ being the swap operator defined by relation
$\ten{V}^{(3)}\ket{\varphi_{1}}\ket{\varphi_{2}}=\ket{\varphi_{2}}\ket{\varphi_{1}}$
for $\ket{\varphi_{i}}\in\mathbb{C}^{3}\;(i=1,2)$. Note that
$\ten{I}_{9}-\ten{Q}^{(3)}$ and $\ten{S}^{(3)}$ are
%where $\ten{I}_{9}-\ten{Q}^{(3)}$ and $\ten{S}^{(3)}$ are
orthogonal projectors and therefore
$\big|X^{T_{B'}}_{3}\big|=(1/11)\big[2\ten{S}^{(3)}+\ten{I}_{9}-\ten{Q}^{(3)}\big]$.
Since
%
%\begin{equation}
$\big|X^{T_{B'}}_{3}\big|^{T_{B'}}=(1/11)\big[3\ten{P}_{+}^{(3)}+\ten{Q}^{(3)}\big]$,
one may easily conclude that
$\big|X^{T_{B'}}_{3}\big|^{T_{B'}}\ge 0$. This with the aid of
the fact that the matrix $X_{3}$ is symmetric and real, ensures
that $\stan$ represents a quantum state.

Now we are in position to prove that $\stan$ satisfies PPT
(positive partial transpose) criterion \cite{Peres,sep}. To this
aim we show that transposition with respect to subsystem $BB'$
preserves the positivity. Indeed, the state $\stan$ transposed
with respect to the $BB'$ subsystem, remains a positive operator.
To see this fact explicitly, we write
\begin{equation}
\stan^{T_{BB'}}=\frac{11}{40} \left[
\begin{array}{cccc}
\left|X_{3}\right|^{T_{B'}} & 0 & 0 & 0\\
0 & \left|X^{T_{B'}}_{3}\right| & X^{T_{B'}}_{3} & 0\\
0 & X^{T_{B'}}_{3} & \left|X^{T_{B'}}_{3}\right| & 0\\
0 & 0 & 0 & \left|X_{3}\right|^{T_{B'}}
\end{array}
\right].
\end{equation}
Positivity of the above operator stems from two facts. As
previously mentioned $|X_{3}|^{T_{B'}}\ge 0$ and on the other hand
the off-diagonal elements are blocked by
$\big|X^{T_{B'}}_{3}\big|$.

\subsubsection{Construction of general $2D \otimes 2D$ case}

Trying to generalize the above investigations we start from the
symmetric matrix
\begin{equation}\label{Xd}
X_{D}=\frac{1}{D^{2}+2D-4}\left[(D-2)\ten{P}_{+}^{(D)}-2\ten{P}^{(D)}+\ten{Q}^{(D)}\right],
\end{equation}
where, as previously, $\ten{P}_{+}^{(D)}$ represents a projector
onto maximally entangled state $\ket{\Psi_{+}^{(D)}}$. Orthogonal
projectors $\ten{P}^{(D)}$ and $\ten{Q}^{(D)}$ are given by
%
%\begin{equation}
%\ten{P}_{+}^{(d)}=\proj{\psi_{+}^{(d)}}, \qquad
%\ket{\psi_{+}^{(d)}}=\frac{1}{\sqrt{d}}\sum_{i=1}^{d}\ket{ii}
%\end{equation}
%
\begin{equation}
%\ten{P}^{(D)}=\frac{1}{D} \left[
%(D-1)\sum_{i=0}^{D-1}\proj{ii}-\sum_{\substack{i,j=0\\i\neq
%j}}^{D-1}\ket{ii}\bra{jj} \right],
\ten{P}^{(D)}=\sum_{i=0}^{D-1}\proj{ii}-\ten{P}^{(D)}_{+}, \qquad
\ten{Q}^{(D)}=\ten{I}_{D^{2 }}-\sum_{i=0}^{D-1}\proj{ii}.
%\left[
%\begin{array}{cccc}
%d-1 & -1 & \ldots & -1\\
%-1 & d-1 & \ldots & -1\\
%\vdots & \vdots & \ddots & \vdots\\
%-1 & -1 & \ldots & d-1
%\end{array}
%\right]
\end{equation}
From Eq. (\ref{Xd}) we have
%\begin{equation}
%\end{equation}
%
\begin{equation}
|X_{D}|=\frac{1}{D^{2}+2D-4}\left[(D-2)\ten{P}_{+}^{(D)}+2\ten{P}^{(D)}+\ten{Q}^{(D)}\right]
\end{equation}
and therefore $|X_{D}|^{T_{B'}}\ge 0$. Subsequently, after
elementary steps, we may obtain
$X_{D}^{T_{B'}}=[1/(D^{2}+2D-4)][2\ten{S}^{(D)}-(\ten{I}_{D^{2}}-\ten{Q}^{(D)})]$
with $\ten{S}^{(D)}$ being defined as
\begin{equation}\label{SD}
\ten{S}^{(D)}=\frac{1}{2}\left[\ten{I}_{D^{2}}+\ten{V}^{(D)}-2\sum_{i=0}^{D-1}\proj{ii}\right],
\end{equation}
where $\ten{V}^{(D)}$ is the swap operator acting on
$\mathbb{C}^{D}\ot\mathbb{C}^{D}$.
%$\ten{I}_{D^{2}}-\ten{Q}^{(D)}$ and $\ten{S}^{(D)}$ being
Again, the projectors $\ten{I}_{D^{2}}-\ten{Q}^{(D)}$ and
$\ten{S}^{(D)}$ are orthogonal and therefore
\begin{equation}
\left|X_{D}^{T_{B'}}\right|=\frac{1}{D^{2}+2D-4}\left[2\ten{S}^{(D)}+\ten{I}_{D^{2}}-\ten{Q}^{(D)}\right].
\end{equation}
Finally, performing partial transposition with respect to
subsystem $B'$, we have
$\big|X_{D}^{T_{B'}}\big|^{T_{B'}}=[1/(D^{2}+2D-4)][D\ten{P}_{+}^{(D)}+\ten{Q}^{(D)}]$
and therefore $\big|X_{D}^{T_{B'}}\big|^{T_{B'}}\ge 0$. Now we
can introduce a class of mixed states
\begin{equation}
\stan^{(D)}=\frac{1}{4}\frac{D^{2}+2D-4}{D^{2}+D-2} \left[
\begin{array}{cccc}
|X_{D}| & 0 & 0 & X_{D}\\
0 & \left|X_{D}^{T_{B'}}\right|^{T_{B'}} & 0 & 0\\
0 & 0 & \left|X_{D}^{T_{B'}}\right|^{T_{B'}} & 0\\
X_{D} & 0 & 0 & |X_{D}|
\end{array}
\right].
\label{1}
\end{equation}
%\begin{equation}
%\left|X_{d}^{T_{B'}}\right|^{T_{B'}}=\left[
%\begin{array}{cccc}
%1 & 0 &
%\end{array}
%\right]
%\end{equation}
%
Again, by virtue of the fact that $|X_{D}|^{T_{B'}}\ge 0$ one may
infer that partial transposition with respect to subsystem $BB'$
preserves the positivity of $\stan^{(D)}$. In other words, the
following matrix
\begin{equation}
\stan^{(D)T_{BB'}}=\frac{1}{4}\frac{D^{2}+2D-4}{D^{2}+D-2} \left[
\begin{array}{cccc}
|X_{D}|^{T_{B'}} & 0 & 0 & 0\\
0 & \left|X_{D}^{T_{B'}}\right| & X_{D}^{T_{B'}} & 0\\
0 & X_{D}^{T_{B'}} & \left|X_{D}^{T_{B'}}\right| & 0\\
0 & 0 & 0 & |X_{D}|^{T_{B'}}
\end{array}
\right].
\label{2}
\end{equation}
possess the nonnegative eigenvalues.

\subsection{Proof of nonzero distillable key $K_{D}$}

Since the state $\stan^{(D)}$ given by Eq. (\ref{1}) is PPT one
might expect that it is separable - it satisfies necessary (PPT)
condition for separability \cite{Peres}. However this is not the
case. This is in agreement with the fact \cite{sep} that PPT
condition is sufficient for separability only for the cases $M
\otimes N$, $MN\leq 6$ while here we have $MN=(2D)^{2}\geq 36$.
Like in \cite{KH2} we shall prove nonseparability of $\stan^{(D)}$
in a very nonstandard way. We simply show that the state has
nonzero $K_{D}$. Such a state must be entangled since, due to
seminal result of Ref. \cite{Norbert}, any separable state has
$K_{D}=0$. Because state is PPT and entangled it must be bound
entangled \cite{bound}.

To prove the cryptographic use of $\stan$, below we show that
there exist a LOCC protocol that allows Alice and Bob to approach
arbitrarily closely to some pbit $\gamma_{ABA'B'}^{(2)}$. Note
that, obviously, since LOCC operations preserves PPT property (see
\cite{bound}), the resulting state is still bound entangled. Given
$k$ copies of the state $\stan$ in the $i$-th step of the protocol
Alice and Bob perform the following operations:
\begin{itemize}
    \item They take the state $\stan^{(i-1)}\;(i=1,\ldots,k-1)$ and one of remaining
    $k-i+1$ copies of $\stan$ (here $\stan^{(0)}=\stan$).
    \item They perform C-NOT operation treating qubits $A$ and $B$
    of $\stan$ as source qubits and that of $\stan^{(i-1)}$ as target
    qubits.
    \item They perform a measurement of target qubits in
    computational basis and then compare their results. If
    both of them have the same results ($00$ or
    $11$) then they keep the source state. Otherwise they get rid of
    it.
\end{itemize}
After performing all $k$ steps, with some probability they arrive
at the following state
\begin{equation}
\stan^{(D,k)}= \frac{1}{N_{D,k}}\left[
\begin{array}{cccc}
|X_{D}|^{\ot k} & 0 & 0 & X_{D}^{\ot k}\\
0 & \left(\left|X_{D}^{T_{B'}}\right|^{T_{B'}}\right)^{\ot k} & 0 & 0\\
0 & 0 & \left(\left|X_{D}^{T_{B'}}\right|^{T_{B'}}\right)^{\ot k} & 0\\
X_{D}^{\ot k} & 0 & 0 & |X_{D}|^{\ot k}
\end{array}
\right],
\end{equation}
where (for $D\ge 3$)
\begin{equation}
N_{D,k}=2{\rm Tr}|X_{D}|^{\ot k}+2{\rm
Tr}\left(\left|X_{D}^{T_{B'}}\right|^{T_{B'}}\right)^{\ot
k}=2\left[1+\left(\frac{D^{2}}{D^{2}+2D-4}\right)^{k}\right]\stackrel{k\to
\infty}{\LRA}2.
\end{equation}
If we define, according to (\ref{Th11}) the matrix,
$\mac{A}_{00,11}^{(A'B')}(D,k)=(1/N_{D,k})X_{D}^{\ot k}$ we can
see that $\big|\big|\mac{A}_{00,11}^{(A'B')}(D,k)\big|\big|_{\Tr}\to 1/2$ whenever
$k\to \infty$, for $D\ge 3$. This means that repeating the whole
procedure described above, one may find such a $k$ that the trace
norm of the upper right block of $\stan^{(D,k)}$ is close to $1/2$
with arbitrary precision. According to the the Theorem 1 (see
Section 2) this convergence guarantees that the original state
$\stan^{(D)}$ defined by the formula (\ref{1}) satisfies
$K_{D}>0$. The Theorem 2 guarantees in particular, that (like it
was in \cite{KH2}) the above sequence of bound entangled states
$\stan^{(D,k)}$ approaches private bit.

\section{Summary and discussion}

We have summarized main elements of general scheme of distillation
of secure key \cite{KH2,KH}. The central notion of the scheme is
the idea of private bit (with its natural generalization - private
dit) which is the state that, in general, consists of two parts:
the key part $AB$ and the shield part $A'B'$. The first contains a
bit of secure key. The role of the second part is - in a sense -
to protect the key form Eve. A surprising fact, found in
\cite{KH2} is that PPT bound entangled states can approach private
bit in trace norm. Since this convergence is a necessary and
sufficient condition to distill secure key form  the original
state, this means that bound entangled states can serve as a
source of distillable key \cite{KH2,KH}.

First bound entangled states with nonzero distillable
key $K_{D}$ were provided in \cite{KH2}. They required
however very high dimensions. The small ($4\otimes 4$)
bound entangled  states with nonzero distillable key
were provided later in paper \cite{KH3}.

Here we have provided new class of  small (of, among others, $6
\otimes 6$ type) bound entangled states with that property. We
have proven this fact applying an easy criterion form \cite{KH2}
showing that a given sequence of quantum states approaches the
sequence of private bits. The LOCC protocol applied to produce
such a sequence was of two-way type. As noticed in \cite{KH2} at
some point the elements of the sequence got one-way distillable
key which can be distilled with help of Devetak-Winter protocol.
The original problem was that such states were of very high
dimensions. Quite surprisingly the low-dimensional bound entangled
states provided in \cite{KH3} represent one-way distillable key.
It was proven with help of the observation that any biased mixture
of two private bits with second of them having the key part
rotated locally with $\sigma_{x}$ Pauli matrix contains nonzero
distillable key. Namely one has the following theorem
\cite{KH3}:\\

\Th{2}{For two private bits $\gamma^{(2)}_{1}$, $\gamma^{(2)}_{2}$
one-way distillable key of the mixture of the form}:
\begin{equation}
\rho= p_{1}\gamma_{1}^{(2)} +
p_{2}\sigma^{(A)}_{x}\gamma_{2}^{(2)}\sigma^{(A)}_{x}
\label{eq:rho-prop}
\end{equation}
{\it with,
%say, $ p_{1} > p_{2} $ and
$ \sigma^{(A)}_{x} = [\sigma_{x}]_{A} \otimes I_{A'BB'}$ satisfies
$K_D^{\rightarrow}(\rho) \geq 1- h(p_1)$ with binary entropy
$h(p_1)$}\footnote{Binary entropy of the distribution $\{ p_{1},
p_{2} \}$ is defined as $h(p_{1})=-p_{1} \log_{2} p_{1} -
p_{2}\log_{2}p_{2}$}.\\
Here $K_{D}^{\rightarrow}(\rho)$ stands for cryptographic key
distillable from $\rho$ with help of forward classical
communication. The natural question is whether one can modify the
$6 \otimes 6$ bound entangled states provided in the present paper
to get mixture of two private bits of the above form. Our first
analysis has shown that most probably it is impossible to turn our
example into a state of the form (\ref{eq:rho-prop}) while keeping
bound entanglement property. In Ref. \cite{KH3} the construction
leading to {\it bound entangled} state of the form
(\ref{eq:rho-prop}) was based on some properties of states that
were used in locking entanglement measures effects. May be that
was the reason why the construction was successful there. Still
there is a natural question about other methods to construct
low-dimensional bound entangled states with one-way distillable
key.

Of course the most important open problem is whether any entangled
bipartite state contains nonzero distillable key or not. In
multipartite case it is not true - there are states that are
entangled but no secure key between any of the parties can be
distilled \cite{RAPH}. In bipartite case lack of such states is
guaranteed (via entanglement distillation approach
\cite{Deutsch96}) only for $2 \otimes 2$ and $2 \otimes 3$  cases,
since, as shown in \cite{97,99}, all the entangled states can be
distilled to singlet form in those cases. For $d \otimes d$ with
$d\geq 4$ it is known that at least some bound entangled (i.e.
nondistillable to singlets) states have nonzero distillable key
\cite{KH3}. No example of $3 \otimes 3$ or $2 \otimes 4$ states
with that property is still known.

\section*{Acknowledgments}
The author thank Maciej Demianowicz and Micha\l{} Horodecki for
fruitful discussions. The work is partially supported by Polish
Ministry of Scientific Research and Information Technology grant
under the (solicited) project no. PBZ-MIN-008/P03/2003 and by EC
grants: RESQ, contract no. IST-2001-37559.
%%% ENDofCHANGE

\end{document}